# Superconductivity with topological surface state in $Sr_xBi_2Se_3$

Zhongheng Liu,[†, ‡, §] Xiong Yao,[†, ‡, §] Jifeng Shao,[†, ‡, §] Ming Zuo,[‡] Li Pi,[†, ‡, ‖] Shun Tan,[‡] Changjin Zhang,[†, ‡, ‖, *] and Yuheng Zhang[†, ‡, ‖]

[†]High Magnetic Field Laboratory, Chinese Academy of Sciences and University of Science and Technology of China, Hefei 230026, People's Republic of China

[‡]Hefei National Laboratory for Physical Sciences at Microscale, University of Science and Technology of China, Hefei 230026, People's Republic of China

[‖]Collaborative Innovation Center of Advanced Microstructures, Nanjing University, Nanjing 210093, China

Supporting Information Placeholder

**ABSTRACT:** By intercalation of alkaline-earth metal Sr in $Bi_2Se_3$, superconductivity with large shielding volume fraction (~91.5% at 0.5 K) has been achieved in $Sr_{0.065}Bi_2Se_3$. The analysis of the Shubnikov-de Hass oscillations confirms the 1/2-shift expected from a Dirac spectrum, giving transport evidence of the existence of surface states. Importantly, the $Sr_xBi_2Se_3$superconductor is stable under air, making the $Sr_xBi_2Se_3$ compound an ideal material base for investigating topological superconductivity.

The theoretical predication and successful experimental realizations of topological insulators have opened an exciting research topic in physics and materials fields.[1-4] These and related materials have attracted much interest not only in investigating their exotic topological properties, but also in the searching of new topological phases. A particularly exciting new phase is topological superconductivity[5,6], which is featured by the existence of gapless surface states at the surfaces of a fully gapped superconductor. Due to the unique electronic structure, the topological superconductors are believed to have great potential applications in fault-tolerant topological quantum computing.

Despite of the importance in both the materials science and potential applications, experimental realizations of topological superconductors have been greatly limited. One way to realize possible topological superconductivity is based on the proximity effects at the interface of topological insulator thin films grown on superconducting substrates.[7-9] Besides the proximity induced superconductivity, the realization of possible topological superconductivity in bulk material could be very important, especially in real application. By intercalation Cu in $Bi_2Se_3$ topological insulator, bulk superconductivity can be achieved in single crystals.[10,11] Recently, tremendous experimental and theoretical efforts have been placed in this material in order both to improve the sample quality and to clarify whether or not this and related compounds are really "topological superconductors".[12-23] However, no consensus has yet been reached. The divergence is mainly due to the relatively low superconducting volume fraction (~50%) of the $Cu_xBi_2Se_3$ samples.[10,11,21] At present, the experimental realization of bulk "topological superconductors" remains a big challenge. The answer to this problem greatly relies on the fabrication of an appropriate material with large superconducting volume fraction and the identification of its surface states.

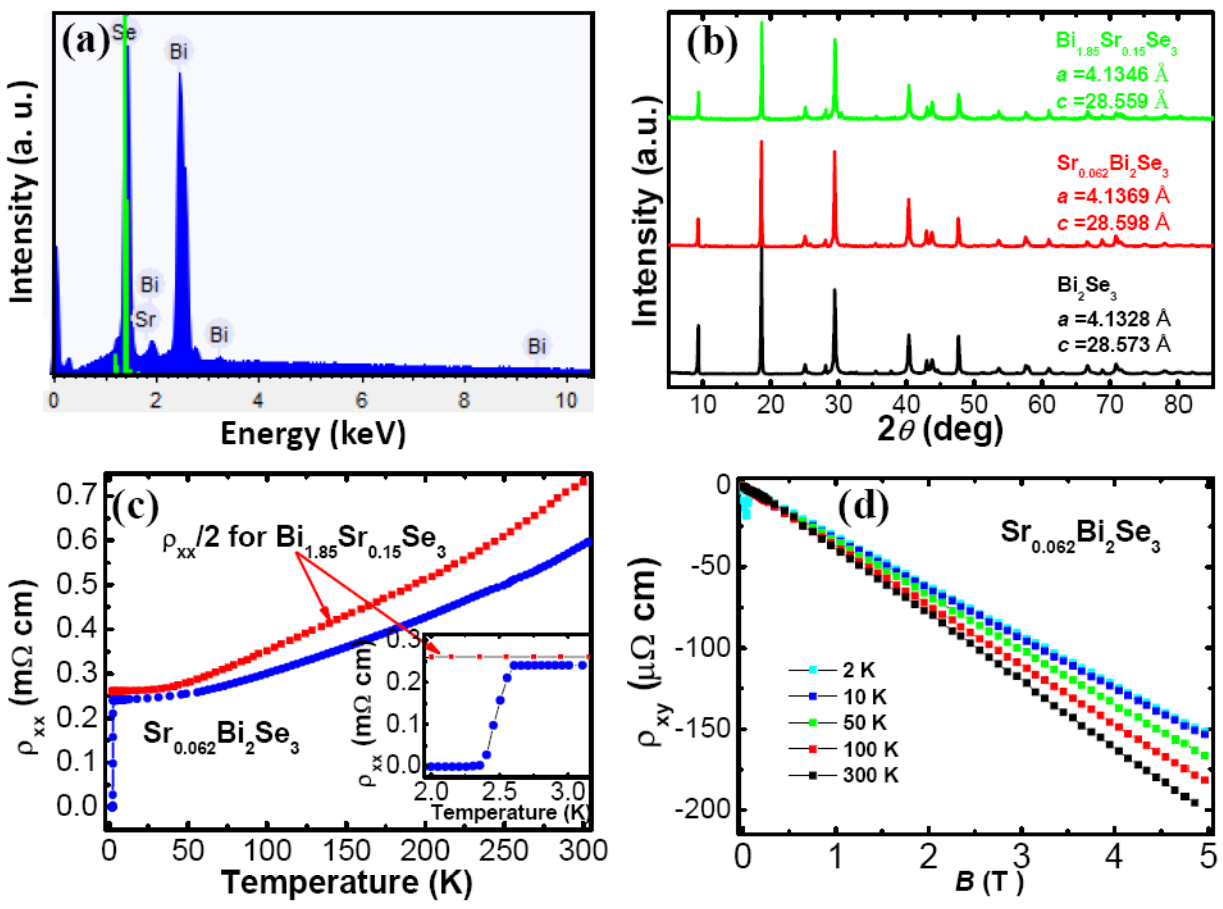

***Figure 1.*** (a) A representative energy dispersive x-ray spectroscopy pattern showing the existence of Sr, Bi, and Se. (b) Powder x-ray diffraction patterns of the $Bi_2Se_3$, $Sr_{0.062}Bi_2Se_3$, $Bi_{1.85}Sr_{0.15}Se_3$ samples. (c) Temperature dependence of resistivity of the $Sr_{0.062}Bi_2Se_3$ sample and the $Bi_{1.85}Sr_{0.15}Se_3$ sample. (d) Magnetic field dependence of transverse resistivity of the $Sr_{0.062}Bi_2Se_3$ sample measured at different temperatures.

Here we show that intercalation of Sr in the well-known topological insulator $Bi_2Se_3$ could lead to a superconducting state below ~2.5 K. The bulk superconductivity has been confirmed by the large shielding volume fraction (91.5%). The quantum oscillations data measured with magnetic field up to 35 Tesla give the -1/2 intercept in the limit $1/B \rightarrow 0$, providing transport evidence of the existence of surface states. Moreover, the $Sr_xBi_2Se_3$ samples are stable in air, which is important in real application.

We have grown a series of $Sr_xBi_2Se_3$ samples with nominal Sr content between 0 and 0.3. The typical dimensions of the obtained single crystals are about 3×3×0.5 $mm^3$. The resultant crystals are easily cleaved along the basal plane leaving a silvery shining mirror like surface. In order to determine the chemical compositions of the obtained samples, we perform an energy dispersive x-ray spectroscopy (EDX) analysis on the $Sr_xBi_2Se_3$ samples. The EDX spectrum shown in Fig. 1a confirms the existence of Sr, Bi, and Se. However, we find that for each sample, the real Sr content is less than the nominal content (Table S1). In particular, the real Sr content can be only about 0.065 when the nominal Sr content is x=0.19. For $Bi_{2-x}Sr_xSe_3$, the actual Sr contents are comparable to the nominal compositions.

Figure 1b shows the powder x-ray diffraction (XRD) patterns of the $Bi_2Se_3$, $Sr_{0.062}Bi_2Se_3$, and $Bi_{1.85}Sr_{0.15}Se_3$ samples. The detailed refinements on the XRD patterns suggest that the lattice parameters of $Sr_xBi_2Se_3$ are a=4.1369 Å and c=28.598 Å, which are larger than those of a=4.1328 Å and c=28.573 Å in undoped $Bi_2Se_3$. And the lattice constants of the $Bi_{1.85}Sr_{0.15}Se_3$ sample is determined to be a=4.146 Å and c=28.559 Å. It is found that the c-axis lattice constant of $Bi_{2-x}Sr_xSe_3$ decreases with increasing Sr doping content. Thus the slight increase in *c*-axis constant in $Sr_{0.062}Bi_2Se_3$ supports that the Sr atoms are intercalated in the $Bi_2Se_3$ lattice.

Figure 1c gives the temperature dependence of in-plane resistivity ($\rho_{xx}$~$T$ curve) of the $Sr_{0.062}Bi_2Se_3$ sample and the $Bi_{1.85}Sr_{0.15}Se_3$ sample. The resistivity of $Sr_{0.062}Bi_2Se_3$ sample exhibits metallic-like behavior at high temperature. Below ~50 K, the $\rho_{xx}$~$T$ curve is very flat, giving the residual resistivity $\rho_{xx0}$=0.24 mΩ cm. The onset of superconducting transition occurs at $T_c$~2.57 K and zero resistivity is achieved at $T_{zero}$~2.39 K. For the $Bi_{1.85}Sr_{0.15}Se_3$ sample, no superconducting transition has been observed down to 1.9 K, suggesting that the Sr-doped $Bi_2Se_3$ samples are not superconducting (we have confirmed that all $Bi_{2-x}Sr_xSe_3$ samples are not superconducting within x≤0.25, see Fig. S6). In order to determine the type and density of charge carriers in the $Sr_{0.062}Bi_2Se_3$ sample, we perform the measurement of Hall resistivity. Figure 1d shows that the Hall resistivity $\rho_{xy}$ is almost proportional to the applied magnetic field *B*, suggesting the dominance of only one type of bulk carrier. The negative slope of $\rho_{xx}$~$B$ curve means that the dominant charge carriers in $Sr_{0.062}Bi_2Se_3$ are electrons. The Hall coefficient $R_H$ slightly decreases with decreasing temperature and the carrier concentration ($n_e$) is found to increase from 2.02×$10^{19}$ $cm^{-3}$ at 300 K to 2.65×$10^{19}$ $cm^{-3}$ below 10 K (Fig. 2a). This carrier density is about six times smaller than that of $Cu_xBi_2Se_3$[11], suggesting that the Fermi level is closer to the Dirac point in the present sample. And the carrier mobility in the $Sr_{0.062}Bi_2Se$ sample is moderate ($\mu$(3K)=1/$n_e\rho_{xx}$(3K)$e$~870 $cm^2$/Vs).

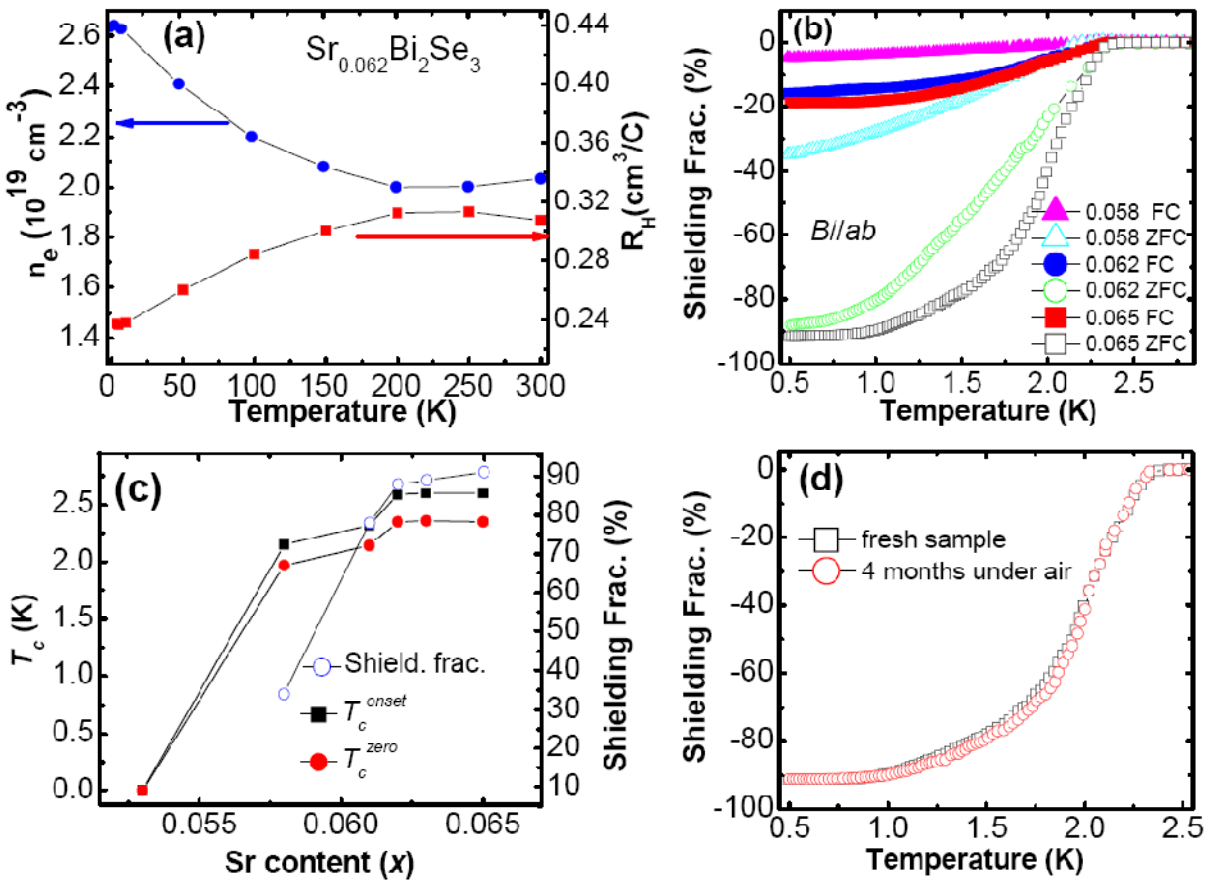

***Figure 2.*** (a) The Hall coefficient $R_H$ and charge carrier density $n_e$ as estimated from Fig. 1d. (b) Temperature dependence of magnetic susceptibility for the $Sr_xBi_2Se_3$ samples. (c) The plot of the onset transition temperature ($T_c^{onset}$), zero-resistivity temperature ($T_c^{zero}$), and the shielding fraction at 0.5 K at different Sr content. (d) A comparison between the temperature dependence of magnetic susceptibility of the fresh-obtained $Sr_{0.065}Bi_2Se_3$ sample and the sample which is exposed under air for 4 months (the same piece).

Figure 2b gives the temperature dependence of magnetic susceptibility for the $Sr_{0.058}Bi_2Se_3$, $Sr_{0.062}Bi_2Se_3$, and $Sr_{0.065}Bi_2Se_3$ samples. The applied magnetic field is 2 Oe with $B||ab$. In the $B||ab$ case, we ignore the effect of demagnetization factor since the dimensions of the samples satisfy $a$~$b \gg c$. For $Sr_{0.058}Bi_2Se_3$, diamagnetic signal appears below 2.2 K, indicating the occurrence of bulk superconductivity. The shielding volume fraction of the $Sr_{0.058}Bi_2Se_3$ sample is about 34% at 0.5 K. For the $Sr_{0.062}Bi_2Se_3$ and $Sr_{0.065}Bi_2Se_3$ samples, diamagnetic signal appears at ~2.35 K. In these samples, large shielding fraction has been obtained. For example, the shielding fraction at 0.5 K can reach to 88% in $Sr_{0.062}Bi_2Se_3$. For $Sr_{0.065}Bi_2Se_3$, the shielding fraction can be as high as 91.5% below 1 K. Figure 2c plots the superconducting transition temperature as well as the shielding fraction (at 0.5 K) for the $Sr_xBi_2Se_3$ samples. Both the transition temperature and the shielding fraction increase with increasing Sr content when x≤0.065. We have placed the $Sr_xBi_2Se_3$ samples under air for four months and found that the superconducting volume fraction does not decrease (Fig. 2d), suggesting that the samples are stable. This is in sharp contrast to that in $Cu_xBi_2Se_3$ samples, where the superconductivity is damaged over several-hours exposure under air.

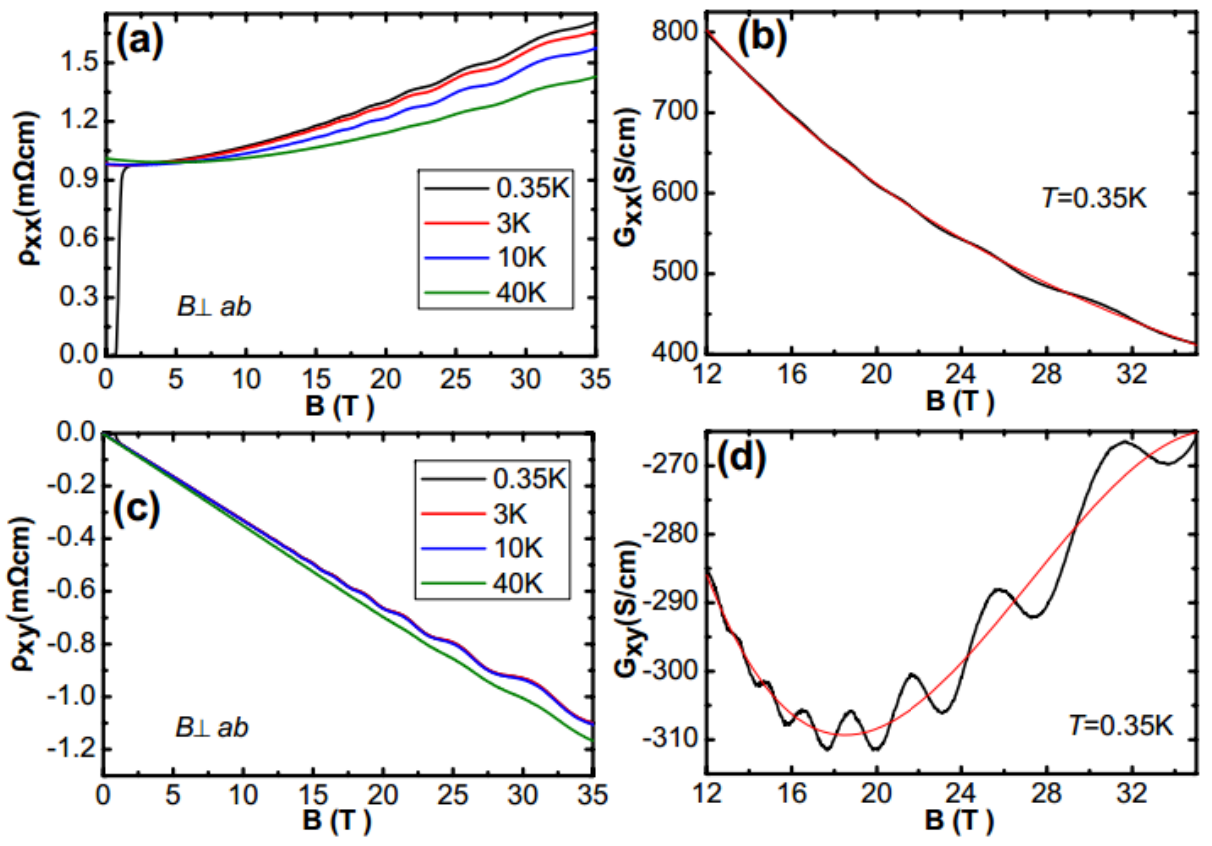

***Figure 3.*** (a) Magnetic field dependence of in-plane resistivity at different temperatures with magnetic field perpendicular to the *ab* plane. (b) The converted in-plane conductance at 0.35 K. (c) Magnetic field dependence of Hall resistivity. (d) The converted Hall conductance at 0.35 K.

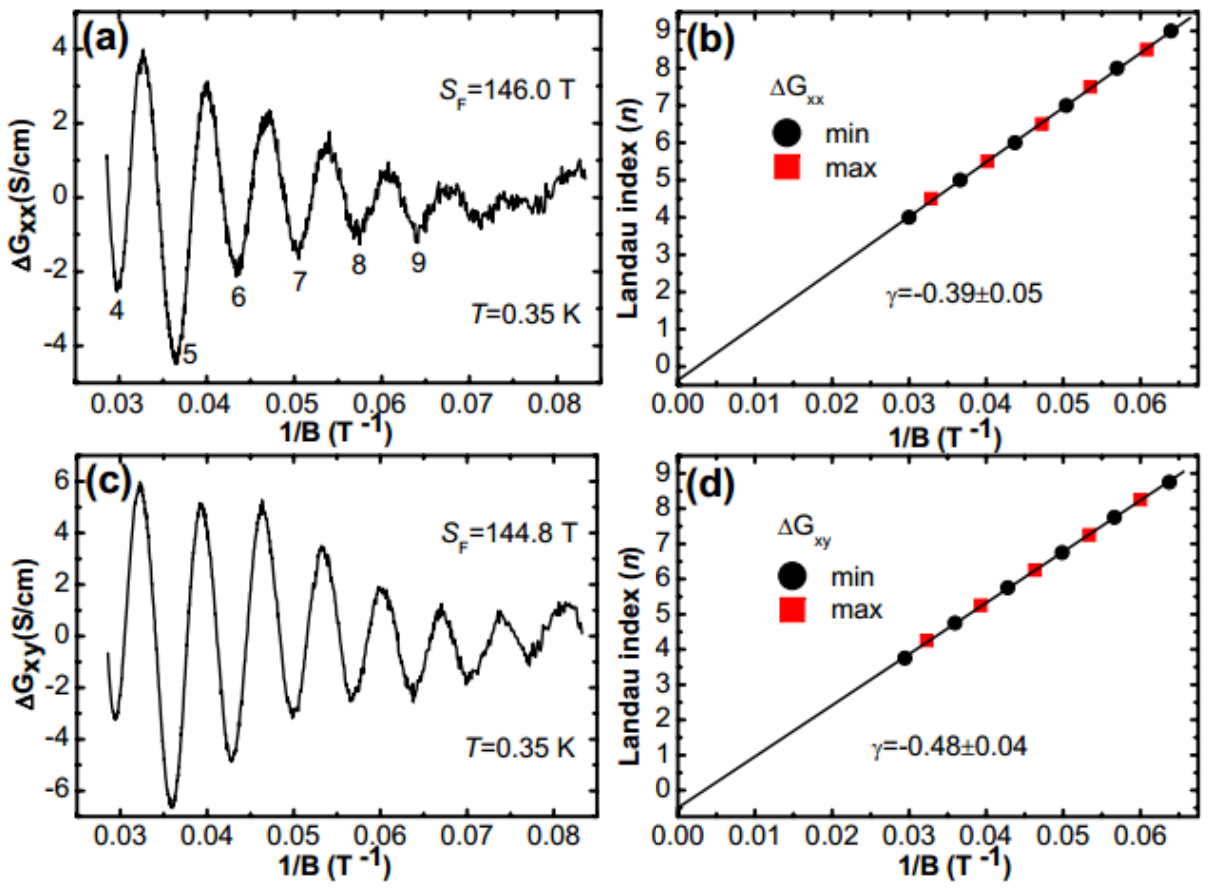

***Figure 4.*** (a) Oscillatory component of the in-plane conductance at 0.35 K plotted against *1/B*. The Landau level indices *n*=4, 5, 6,... are indicated for the minima of $\Delta G_{xx}$. (b) Landau index (*n*) vs *1/B*, where *n* and *n+1/2* correspond to the minima and maxima of $\Delta G_{xx}$, respectively. (c) Oscillatory component of the Hall conductance at 0.35 K plotted against *1/B*. (d) Landau index vs *1/B* derived from (c), where *1/B* is plotted against *n+1/4*. The *1/4* shift arises because the minima in $d\Delta G_{xy}/dB$ align with the minima in $\Delta G_{xx}$.[25]

In order to determine whether or not there is a surface state in the $Sr_xBi_2Se_3$ system, we perform the quantum oscillation measurements on the $Sr_{0.062}Bi_2Se_3$ sample. Figures 3a and c give the magnetic field dependence of in-plane resistivity ($\rho_{xx}$) and Hall resistivity ($\rho_{xy}$) of the $Sr_{0.062}Bi_2Se_3$ sample measured with $B \perp ab$, respectively. At 0.35 K, the pronounced Shubnikov-de Haas (SdH) oscillation can be observed when the magnetic field is larger than 10 T both in the $\rho_{xx}$~*B* curve and in the $\rho_{xy}$~*B* curve. With increasing temperature, the amplitude of the SdH oscillation decreases. When *T*>40 K, the magnitudes of the SdH oscillations are greatly weakened. From the $\rho_{xy}$~*B* curves in Fig. 3c it is estimated that the electron carrier concentration is about $n_e$~$2.6\times10^{19}$ cm$^{-3}$, consistent with that given in Fig. 2a.

In bulk $Bi_2Se_3$-type samples, the surface conductance ($G^s$) is usually small comparing to the bulk conductance ($G^b$).[24,25] When the surface conduction channel coexists with the bulk conduction channel, the observed conductance matrix is the sum $G_{xx}=G_{xx}^{s}+G_{xx}^{b}$ and $G_{xy}=G_{xy}^{s}+G_{xy}^{b}$. In this case, the experimentally measured resistivity and Hall effect are contributed from both the surface conductance ($G_{xx}^{s}$ and $G_{xy}^{s}$) and the bulk conductance ($G_{xx}^{b}$ and $G_{xy}^{b}$). Thus simply from the measured $\rho_{xx}$~*B* curves and the $\rho_{xy}$~*B* curves one cannot determine the Landau index. As has been suggested by Xiong *et al.*[25], it is expedient to convert the $\rho_{xx}$ and $\rho_{xy}$ into the conductance $G_{xx}=\rho_{xx}/[\rho^2_{xx}+\rho^2_{xy}]$ and $G_{xy}=\rho_{xy}/[\rho^2_{xx}+\rho^2_{xy}]$. In Figs. 3b and d we plot the converted $G_{xx}$~*B* curve and the $G_{xy}$~*B* curve at 0.35 K, respectively. Here we show only the data at 0.35 K because the $G_{xx}$~*B* curves and the $G_{xy}$~*B* curves at other temperature exhibit the same oscillatory period with the curves at 0.35 K.

Figure 4a shows the SdH oscillation of $\Delta G_{xx}$ against 1/*B* after subtracting the background. The simple oscillatory pattern is the result of the single frequency F=146 T. This frequency is slightly larger than (or comparable to) that in $Bi_2Se_3$[26,27], while it is smaller than that in $Cu_xBi_2Se_3$.[16,17] Figure 4b plots the minima of $\Delta G_{xx}$ (solid circles) and the maxima of $\Delta G_{xx}$ which is shifted by 1/2(red squares). From Fig. 4b one can see that the best-fit straight line intercepts the *n* axis at the value γ=-0.39±0.05. The same analysis is applied to the Hall conductance (Fig. 4c). The frequency determined from the $\Delta G_{xy}$~1/*B* curve is F=144.8 T, in agreement with that determined from the $\Delta G_{xx}$~1/*B* curve. Figure 4d shows the plot based on the minima and maxima of the Hall conductance $\Delta G_{xy}$. The minima in $\Delta G_{xy}$ correspond to *n*+1/4, since the derivative $d\Delta G_{xy}/dB$ has minima at *n*.[25] The intercept of the fitted line occurs at γ=-0.48±0.04. It is found that these intercepts are quite closer to the ideal Dirac case γ=-1/2, rather than the Schrödinger case where γ=0. The slight deviation of γ value from -1/2 could be probably due to an additional phase shift resulting from the curvature of the Fermi surface in the third direction, which changes from 0 for a quasi-2*D* cylindrical Fermi surface to ±1/8 for a corrugated 3*D* Fermi surface (-0.39±0.05=-(0.515-1/8) ±0.05).[28] Thus the high-field Shubnikov-de Haas oscillation results clearly reveals the existence of a nontrival π Berry's phase (γ=1/2), and thus provides strong evidence of the existence of Dirac fermions in $Sr_xBi_2Se_3$. In topological insulators $Bi_2Se_3$ and $Bi_2Te_3$, the gapless Dirac fermions are accompanied with the surface states. The present Shubnikov-de Hass oscillation data may provide transport evidence for the existence of surface states in the $Sr_xBi_2Se_3$samples. The existence of topological properties in $Sr_xBi_2Se_3$ superconductors could be further confirmed by angle-resolved photoemission spectroscopy, scanning tunneling spectroscopy, as well as other experiments.

In conclusion, we have successfully grown a series of high-quality $Sr_xBi_2Se_3$ single crystal superconductors. The samples exhibit very high shielding volume fraction (91.5%). From the high-field Shubnikov-de Hass oscillations we find the transport evidence of the existence of the surface states, probably suggesting that the $Sr_xBi_2Se_3$ compounds are "topological superconductors". More importantly, the $Sr_xBi_2Se_3$ samples are completely insensitive to air. The stability of the $Sr_xBi_2Se_3$ samples is very helpful both in future fundamental investigation and in possible electronic applications.

## ASSOCIATED CONTENT

### Supporting Information

Experimental details, supporting table and figures are included in the supporting information. This material is available free of charge via the Internet at http://pubs.acs.org.

## AUTHOR INFORMATION

### Corresponding Author

zcjin@ustc.edu.cn

### Author Contributions

[§]Z.L., X.Y. and J.S. contributed equally to this work

Notes

The authors declare no competing financial interest.

## ACKNOWLEDGMENT

We acknowledge Xianhui Chen, Zhong Fang, Donglai Feng, and Xingjiang Zhou for helpful discussions. This work was supported by the National Natural Science Foundation of China (Grant Nos. 11174290 and U1232142), and the Scientific Research Grant of Hefei Science Center of Chinese Academy of Sciences (Grant No. 2015SRG-HSC025).

# Superconductivity with topological surface state in $Sr_xBi_2Se_3$

*Zhongheng Liu,[†,‡] Xiong Yao,[†,‡] Jifeng Shao,[†,‡] Ming Zuo,[‡] Li Pi,[†,‡,∥] Shun Tan,[‡] Changjin Zhang,[†,‡,∥,*] and Yuheng Zhang[†,‡,∥]*

[†] *High Magnetic Field Laboratory, Chinese Academy of Sciences and University of Science and Technology of China, Hefei 230026, People's Republic of China*

[‡] *Hefei National Laboratory for Physical Sciences at Microscale, University of Science and Technology of China, Hefei 230026, People's Republic of China*

[∥] *Collaborative Innovation Center of Advanced Microstructures, Nanjing, 210093, China*

*Index Page*

**Sample preparations and experimental details**

**Sample preparations**. Single crystals of $Sr_xBi_2Se_3$as well as$Bi_{2-x}Sr_xSe_3$ were grown by melting stoichiometric mixtures of Bi powder, Sr piece, and Se powder in sealed evacuated quartz glass tubes at 850°C for 48 h, followed by a slow cooling to 610 at a rate of 3°C/h. After that, the quartz tubes were taken out and the samples were quenched down into water ice.

**Scanning electron microscopy and Energy dispersive x-ray spectroscopy.** Scanning electron microscopy measurement was performed using aHitachiTM3000STM. All samples are cylindrical in shape. The energy dispersive x-ray spectroscopy (EDX) analysis was performed using Oxford SWIFT3000 spectroscopy equipped with a Si detector. The EDX measurements were done on single crystals.

**High resolution tunneling electron microscopy and atomic resolution tunneling electron microscopy.** High resolution tunneling electron microscopy (HRTEM) measurement was performed using aJEOL-2010 transmission electron microscope. The point-to-point resolution is better than 0.2 nm. Prior to the HRTEM measurement, the $Sr_xBi_2Se_3$ samples were ground into fine powder specimens. The specimens were then loaded into a copper grid which serves as the sample holder during the HRTEM measurement. The applied accelerating voltage is 200 kV in the measurement. Both the HRTEM images of the specimens and the electron-diffraction patterns were taken. Atomic resolution tunneling electron microscopy measurement was performed using a JEOL-ARM-200F microscope which offers resolution of 0.08 nm at 200 kV. The operational procedure is similar to the HRTEM measurement.

**Shubnikov-de Hass oscillation measurements under high magnetic field**. The magnetic field dependence of in-plane resistivity ($\rho_{xx}$) and Hall resistivity ($\rho_{xy}$) measurements were performed in a thin sample with dimensions of 3×1×0.05 $mm^3$. The $\rho_{xx}$ and $\rho_{xy}$ measurements were performed simultaneously. The schematic illustration of the electrical probes is shown below.

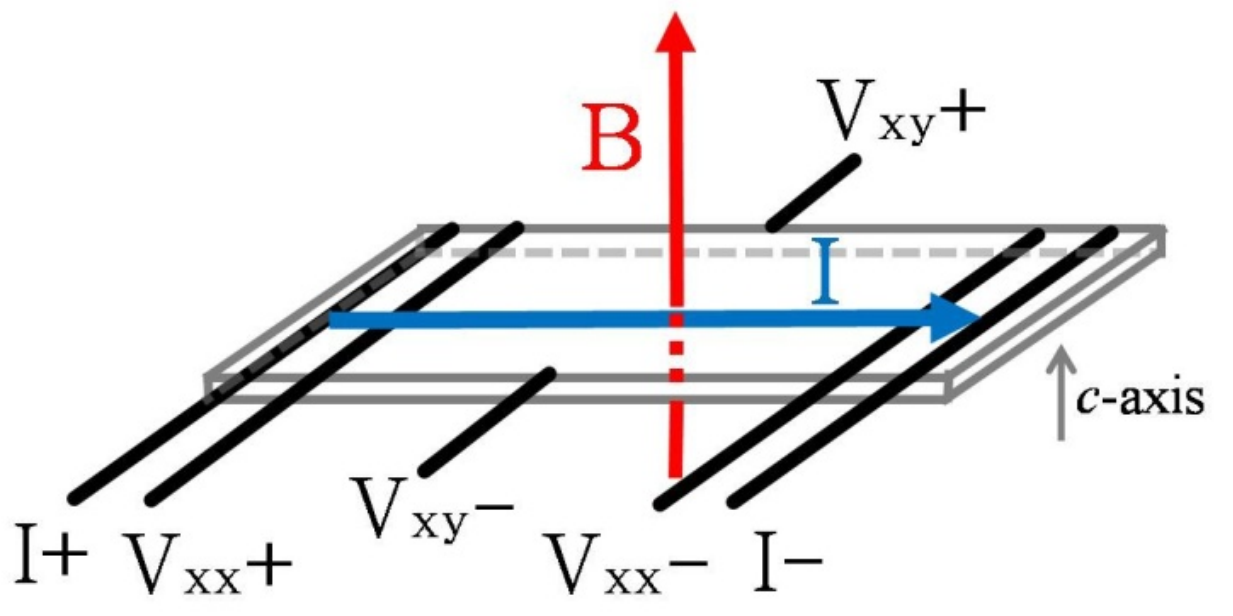

Figure S1. **A schematic illustration of the electrical probes used in the Shubnikov-de Hass oscillation measurements.** The magnetic field dependence of in-plane resistivity and Hall resistivity measurements were performed simultaneously.

In order to determine the chemical compositions of the samples, we perform an energy dispersive x-ray spectroscopy (EDX) analysis by randomly select twenty points. And the average is adopted as the real composition. When the nominal Sr content is higher than x=0.14, the real Sr content in the obtained single crystals exhibits very little change.

**TableS1. A comparison between the nominal composition and the real composition of the $Sr_xBi_2Se_3$ samples. We also give the onset of superconducting transition ($T_c^{onset}$) and the shielding fraction at 0.5 K for comparison.**

| Nominal composition | Real composition | $T_c^{onset}$ (K) | Shield. Frac. at 0.5 K (%) |
|---|---|---|---|
| $Bi_2Se_3$ | $Bi_{2.02}Se_3$ | - | - |
| $Sr_{0.05}Bi_2Se_3$ | $Sr_{0.033}Bi_{2.02}Se_3$ | - | - |
| $Sr_{0.1}Bi_2Se_3$ | $Sr_{0.046}Bi_{2.02}Se_3$ | - | - |
| $Sr_{0.12}Bi_2Se_3$ | $Sr_{0.053}Bi_{2.01}Se_3$ | - | - |
| $Sr_{0.13}Bi_2Se_3$ | $Sr_{0.058}Bi_{2.01}Se_3$ | 2.15 | 34 |
| $Sr_{0.14}Bi_2Se_3$ | $Sr_{0.061}Bi_{2.01}Se_3$ | 2.32 | 78 |
| $Sr_{0.15}Bi_2Se_3$ | $Sr_{0.062}Bi_{2.01}Se_3$ | 2.59 | 88 |
| $Sr_{0.16}Bi_2Se_3$ | $Sr_{0.063}Bi_{2.01}Se_3$ | 2.58 | 87 |
| $Sr_{0.17}Bi_2Se_3$ | $Sr_{0.062}Bi_{2.01}Se_3$ | 2.6 | 88 |
| $Sr_{0.18}Bi_2Se_3$ | $Sr_{0.063}Bi_2Se_3$ | 2.6 | 90 |
| $Sr_{0.19}Bi_2Se_3$ | $Sr_{0.065}Bi_2Se_3$ | 2.6 | 91.5 |
| $Sr_{0.20}Bi_2Se_3$ | $Sr_{0.064}Bi_{2.01}Se_3$ | 2.56 | 88 |
| $Sr_{0.21}Bi_2Se_3$ | $Sr_{0.064}Bi_{2.01}Se_3$ | 2.51 | 84 |
| $Sr_{0.25}Bi_2Se_3$ | $Sr_{0.062}Bi_{1.99}Se_3$ | 2.32 | 69 |
| $Sr_{0.28}Bi_2Se_3$ | $Sr_{0.062}Bi_2Se_3$ | 2.34 | 51 |
| $Sr_{0.3}Bi_2Se_3$ | $Sr_{0.063}Bi_{1.99}Se_3$ | 2.31 | 33 |

**TableS2. The EDX analysis of the sample with nominal composition $Sr_{0.15}Bi_2Se_3$. The real composition determined from EDX analysis is $Sr_{0.062}Bi_{2.01}Se_3$.**

| Point | Se atomic ratio (%) | Sr atomic ratio (%) | Bi atomic ratio (%) | Chemical composition |
|---|---|---|---|---|
| 1 | 59.46 | 1.27 | 39.27 | $Sr_{0.064}Bi_{1.98}Se_3$ |
| 2 | 58.97 | 0.94 | 40.09 | $Sr_{0.048}Bi_{2.04}Se_3$ |
| 3 | 59.22 | 1.26 | 39.52 | $Sr_{0.064}Bi_2Se_3$ |
| 4 | 59.03 | 1.52 | 39.45 | $Sr_{0.077}Bi_{2.01}Se_3$ |
| 5 | 58.91 | 1.17 | 39.92 | $Sr_{0.06}Bi_{2.03}Se_3$ |
| 6 | 58.99 | 1.22 | 39.79 | $Sr_{0.062}Bi_{2.02}Se_3$ |
| 7 | 59.16 | 1.28 | 39.56 | $Sr_{0.065}Bi_{2.01}Se_3$ |
| 8 | 60.04 | 0.91 | 39.05 | $Sr_{0.045}Bi_{1.95}Se_3$ |
| 9 | 59.07 | 1.42 | 39.51 | $Sr_{0.072}Bi_{2.01}Se_3$ |
| 10 | 59.06 | 1.22 | 39.72 | $Sr_{0.062}Bi_{2.02}Se_3$ |
| 11 | 59.21 | 1.28 | 39.51 | $Sr_{0.065}Bi_2Se_3$ |
| 12 | 58.96 | 1.23 | 39.81 | $Sr_{0.063}Bi_{2.03}Se_3$ |
| 13 | 59.62 | 0.77 | 39.61 | $Sr_{0.039}Bi_{1.99}Se_3$ |
| 14 | 59.21 | 1.21 | 39.58 | $Sr_{0.061}Bi_{2.01}Se_3$ |
| 15 | 59.18 | 1.13 | 39.69 | $Sr_{0.057}Bi_{2.01}Se_3$ |
| 16 | 58.78 | 1.31 | 39.91 | $Sr_{0.067}Bi_{2.04}Se_3$ |
| 17 | 58.81 | 1.84 | 39.35 | $Sr_{0.093}Bi_{2.01}Se_3$ |
| 18 | 59.55 | 1.29 | 39.16 | $Sr_{0.065}Bi_{1.97}Se_3$ |
| 19 | 59.31 | 1.22 | 39.47 | $Sr_{0.062}Bi_2Se_3$ |
| 20 | 59.04 | 1.19 | 39.77 | $Sr_{0.06}Bi_{2.02}Se_3$ |
| **Average** | **59.18** | **1.23** | **39.59** | $\mathbf{Sr_{0.062}Bi_{2.01}Se_3}$ |

For the sample with nominal composition $Sr_{0.15}Bi_2Se_3$, the EDX results give an average composition as $Sr_{0.065}Bi_2Se_3$. And the Sr atoms are not very homogenously distributed. In fact, we detect some areas where the Sr contents (less than 0.04) are much lower than the average Sr content (0.062).

**TableS3. The EDX analysis results of the sample with nominal composition $Sr_{0.19}Bi_2Se_3$. The real composition determined from EDX analysis is $Sr_{0.065}Bi_2Se_3$.**

| Point | Se atomic ratio (%) | Sr atomic ratio (%) | Bi atomic ratio (%) | Chemical composition |
|---|---|---|---|---|
| 1 | 59.07 | 1.27 | 39.66 | $Sr_{0.064}Bi_{2.01}Se_3$ |
| 2 | 58.91 | 1.24 | 39.85 | $Sr_{0.063}Bi_{2.03}Se_3$ |
| 3 | 59.29 | 1.21 | 39.5 | $Sr_{0.061}Bi_2Se_3$ |
| 4 | 59.19 | 1.26 | 39.55 | $Sr_{0.064}Bi_2Se_3$ |
| 5 | 59.11 | 1.27 | 39.62 | $Sr_{0.064}Bi_{2.01}Se_3$ |
| 6 | 59.26 | 1.37 | 39.37 | $Sr_{0.069}Bi_{1.99}Se_3$ |
| 7 | 59.3 | 1.19 | 39.51 | $Sr_{0.06}Bi_2Se_3$ |
| 8 | 59.31 | 1.43 | 39.26 | $Sr_{0.072}Bi_{1.99}Se_3$ |
| 9 | 59.32 | 1.16 | 39.52 | $Sr_{0.059}Bi_2Se_3$ |
| 10 | 59.36 | 1.33 | 39.31 | $Sr_{0.067}Bi_{1.99}Se_3$ |
| 11 | 58.98 | 1.39 | 39.63 | $Sr_{0.071}Bi_{2.02}Se_3$ |
| 12 | 59.17 | 1.28 | 39.55 | $Sr_{0.065}Bi_{2.01}Se_3$ |
| 13 | 59.51 | 1.30 | 39.19 | $Sr_{0.066}Bi_{1.98}Se_3$ |
| 14 | 59.23 | 1.26 | 39.51 | $Sr_{0.064}Bi_2Se_3$ |
| 15 | 59.18 | 1.21 | 39.61 | $Sr_{0.061}Bi_{2.01}Se_3$ |
| 16 | 59.24 | 1.23 | 39.53 | $Sr_{0.062}Bi_2Se_3$ |
| 17 | 59.15 | 1.11 | 39.74 | $Sr_{0.056}Bi_{2.02}Se_3$ |
| 18 | 59.17 | 1.66 | 39.17 | $Sr_{0.084}Bi_{1.99}Se_3$ |
| 19 | 59.13 | 1.28 | 39.59 | $Sr_{0.065}Bi_{2.01}Se_3$ |
| 20 | 59.4 | 1.33 | 39.27 | $Sr_{0.067}Bi_{1.98}Se_3$ |
| **Average** | **59.23** | **1.28** | **39.49** | $\mathbf{Sr_{0.065}Bi_2Se_3}$ |

For the sample with nominal composition $Sr_{0.19}Bi_2Se_3$, the EDX results suggest that the distribution of Sr is more homogenous, giving the real composition as $Sr_{0.065}Bi_2Se_3$.

Supplementary Figures

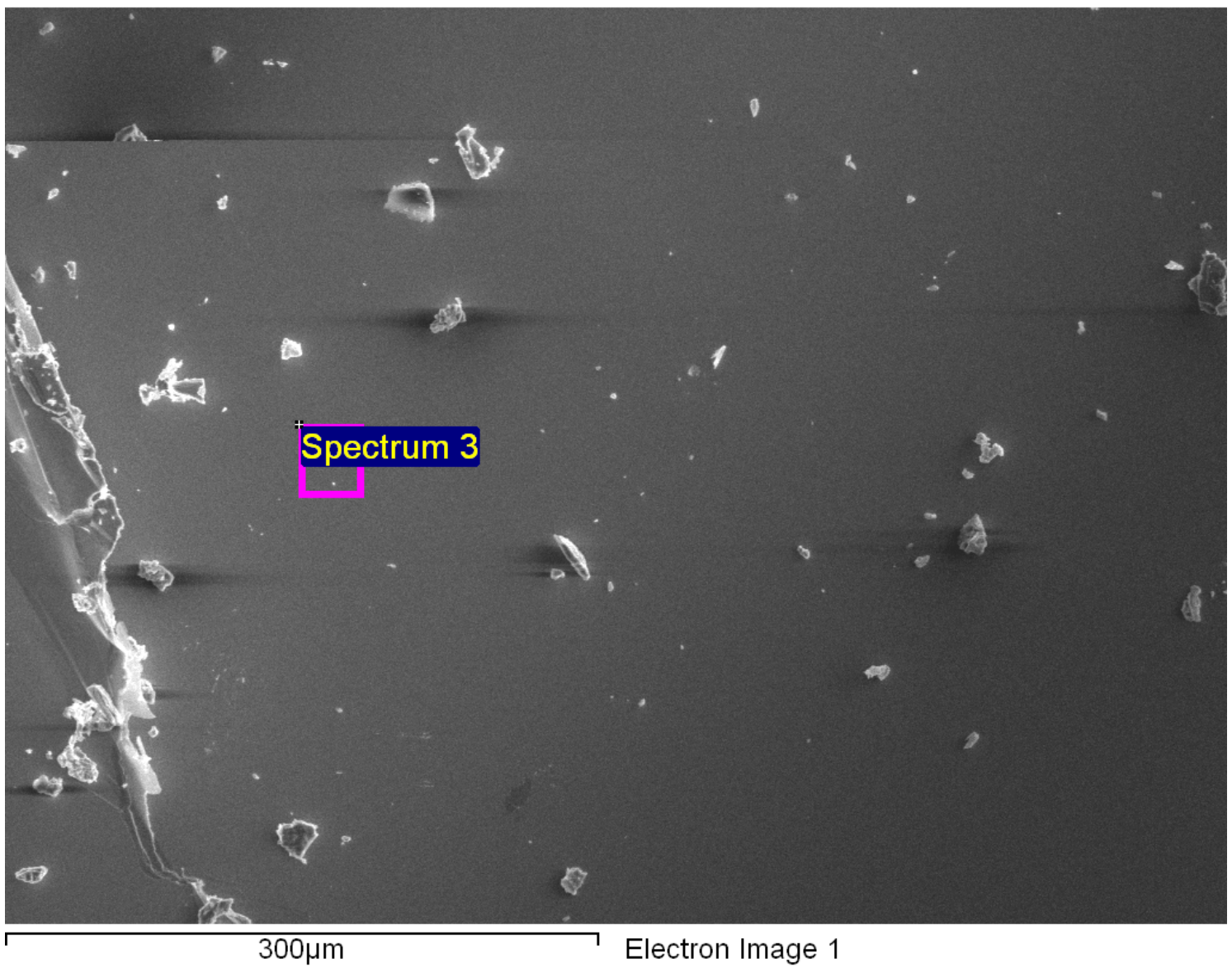

**Figure S2.** **$r_{0.06}Bi_2Se_3$ single crystal sample.**
The surface is flat and shining with silvery color.

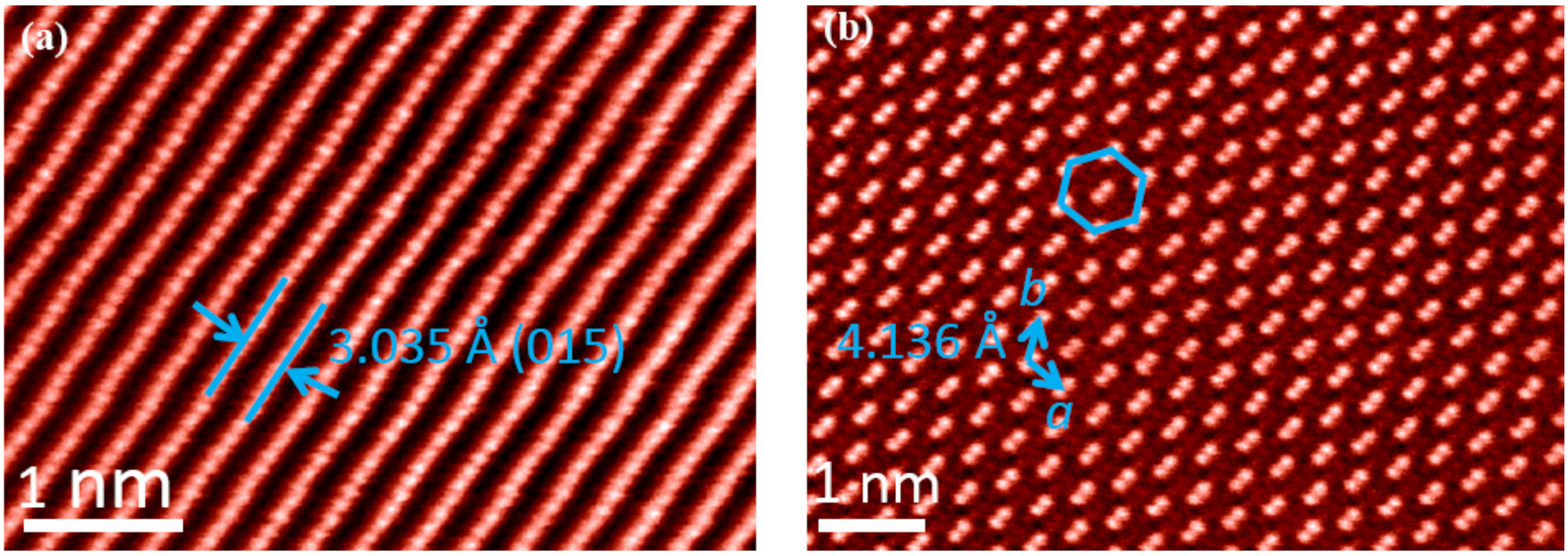

**Figure S3. High resolution transmission electron microscopy pattern of the $Sr_{0.06}Bi_2Se_3$ sample.**

a) High-resolution transmission electron microscopy image of the sample. It shows clearly the equally spaced lattice fringes. The calculated fringe separation is 3.035 Å, which corresponds to the *d*-spacing of the (015) plane of rhombohedral $Sr_xBi_2Se_3$. b) Atomic-resolution transmission electron microscopy image of the sample taken along the [001] zone-axis direction. The rhombohedral arrayed atoms are clearly seen without any stacking defects, suggesting the high quality of the single crystal samples.

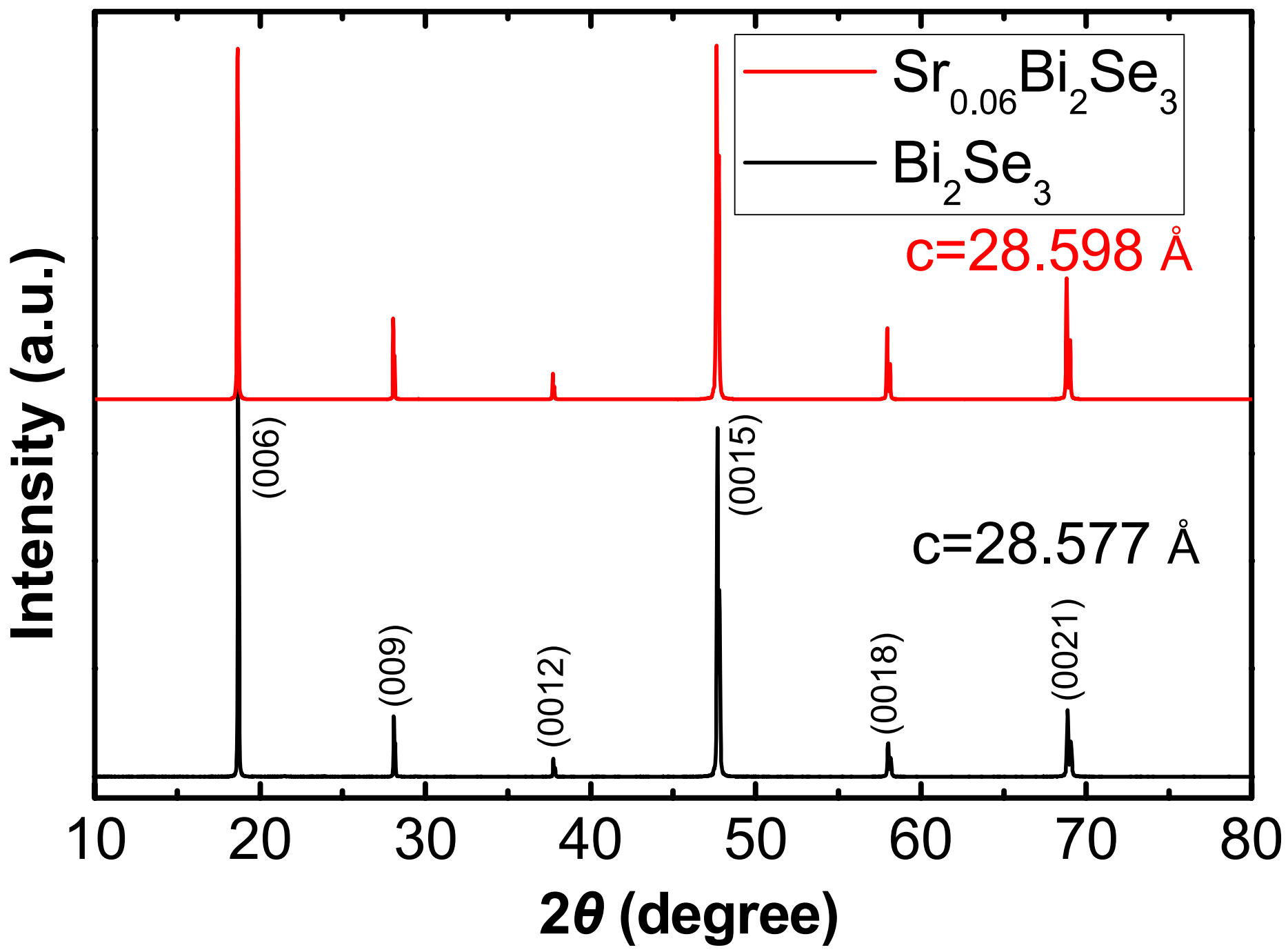

**Figure S4. Single crystal x-ray diffraction pattern of the $Sr_{0.06}Bi_2Se_3$ sample together with the $Bi_2Se_3$ reference sample.**

The single crystal x-ray diffraction data suggest that the cleaved surface is *ab* plane. The full width at half maximum is less than $0.1^{\circ}$ for all the peaks, indicating very high quality of the single crystals. The *c*-axis lattice parameter increases from 28.577 Å for $Bi_2Se_3$ to 28.598 Å for $Sr_{0.06}Bi_2Se_3$, which are consistent with those determined from powder x-ray diffraction data.

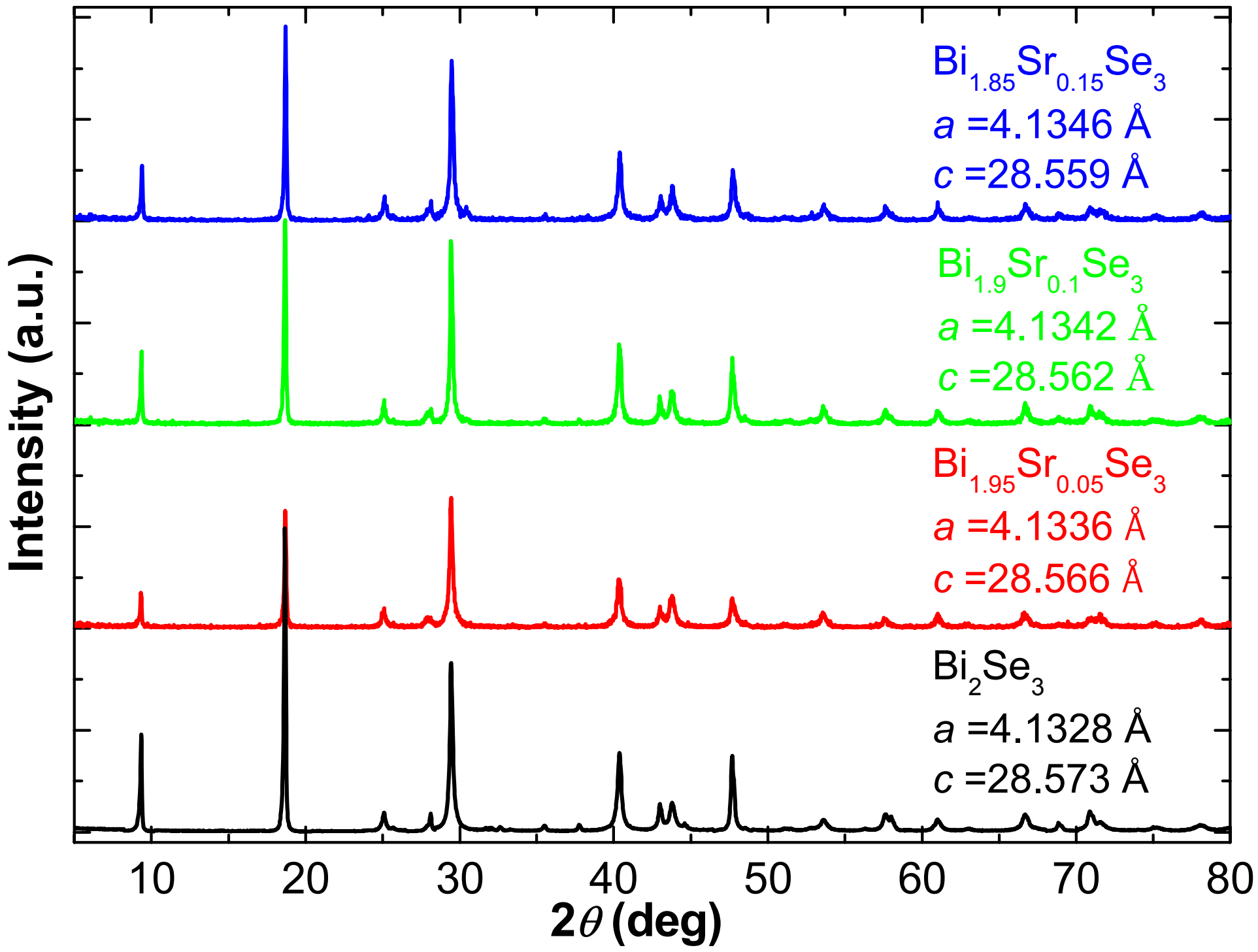

**Figure S5. Powder x-ray diffraction patterns of the $Bi_{2-x}Sr_xSe_3$ samples.**

The *a*-axis lattice constant slightly increases with increasing Sr doping while the *c*-axis lattice constant exhibits slight shrinkage with increasing Sr doping.

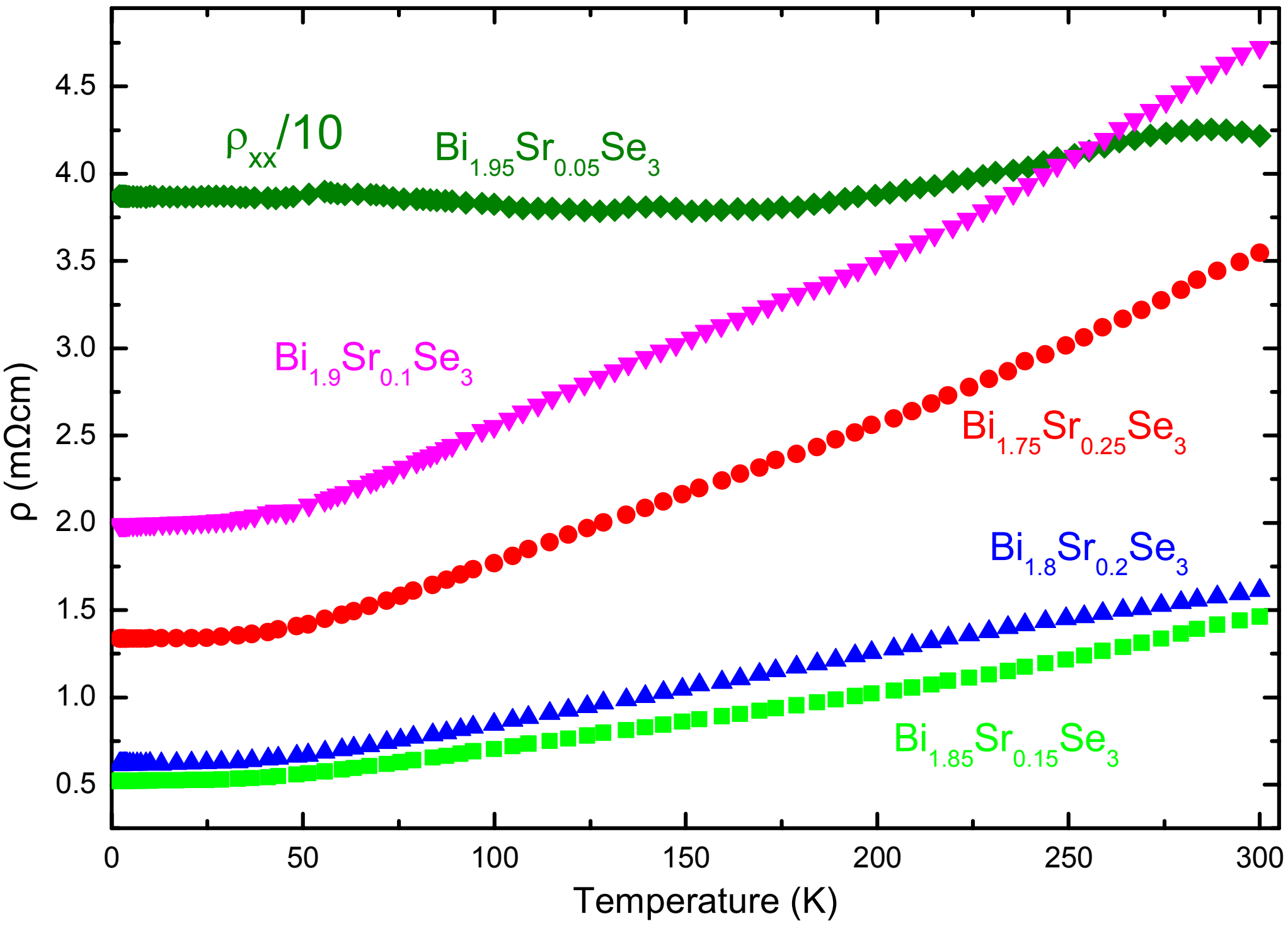

**Figure S6. Temperature dependence of resistivity of the $Bi_{2-x}Sr_xSe_3$ samples.**

It can be seen that for all Sr doped samples, it exhibits no superconducting transition down to 1.9 K.